\pdfoutput=1

\documentclass{appolb}
\usepackage{amssymb,amsmath,graphicx}
\usepackage[T1]{fontenc}

\begin{document}
\title{
  Adiabatic quantum pumping in buckled graphene nanoribbon driven by a~kink 
  \thanks{{\em Presented at 45.\ Zjazd Fizyk\'{o}w Polskich, Krak\'{o}w 13--18
    wrze\'{s}nia 2019.}}
}
\author{Dominik Suszalski and Adam Rycerz
\address{Institute of Theoretical Physics, Jagiellonian University,
  ul.\ S.\ {\L}ojasiewicza~11, PL--30348~Krak\'{o}w, Poland}
}
\date{February 18, 2020}
\maketitle
\begin{abstract}
We propose a new type of quantum pump in buckled graphene nanoribbon,
adiabatically driven by a~kink moving along the ribbon.
From a practical point of view, pumps with moving scatterers present
advantages as compared to gate-driven pumps, like enhanced charge transfer
per cycle per channel. The kink geometry is simplified by truncating
the spatial arrangement of carbon atoms with the classical $\phi^4$ model
solution, including a~width renormalization following from the
Su-Schrieffer-Heeger model for carbon nanostructures.
We demonstrate numerically, as a~proof of concept, 
that for moderate deformations a~stationary kink at the ribbon center
may block the current driven by the external voltage bias. 
In the absence of a~bias, a~moving kink leads to
highly-effective pump effect, with a~charge per unit cycle dependent
on the gate voltage. 
\end{abstract}

\section{Introduction}

The idea of quantum pumping, i.e., transferring the charge between
electronic reservoirs by periodic modulation of the device connecting
these reservoirs \cite{Naz09}, has been widely discussed in the context
of graphene nanostructures \cite{Pra10,Wak10,San12,Jia13,Gri13,Abd17,Fuj20}. 
Since early works, elaborating the gate-driven pumping mechanism in graphene
\cite{Pra10} and bilayer graphene \cite{Wak10}, it becomes clear that
the transport via evanescent modes may significantly enhance the effectiveness
of graphene-based pumps compared to other quantum pumps.
Other pumping mechanisms considered involves laser irradiation \cite{San12},
strain fields \cite{Jia13}, tunable magnetic barriers \cite{Gri13},
Landau quantization \cite{Abd17}, or even sliding the Moir\'{e} pattern
in twisted bilayer graphene \cite{Fuj20}.

\begin{figure}[!t]
\centerline{\includegraphics[width=12.5cm]{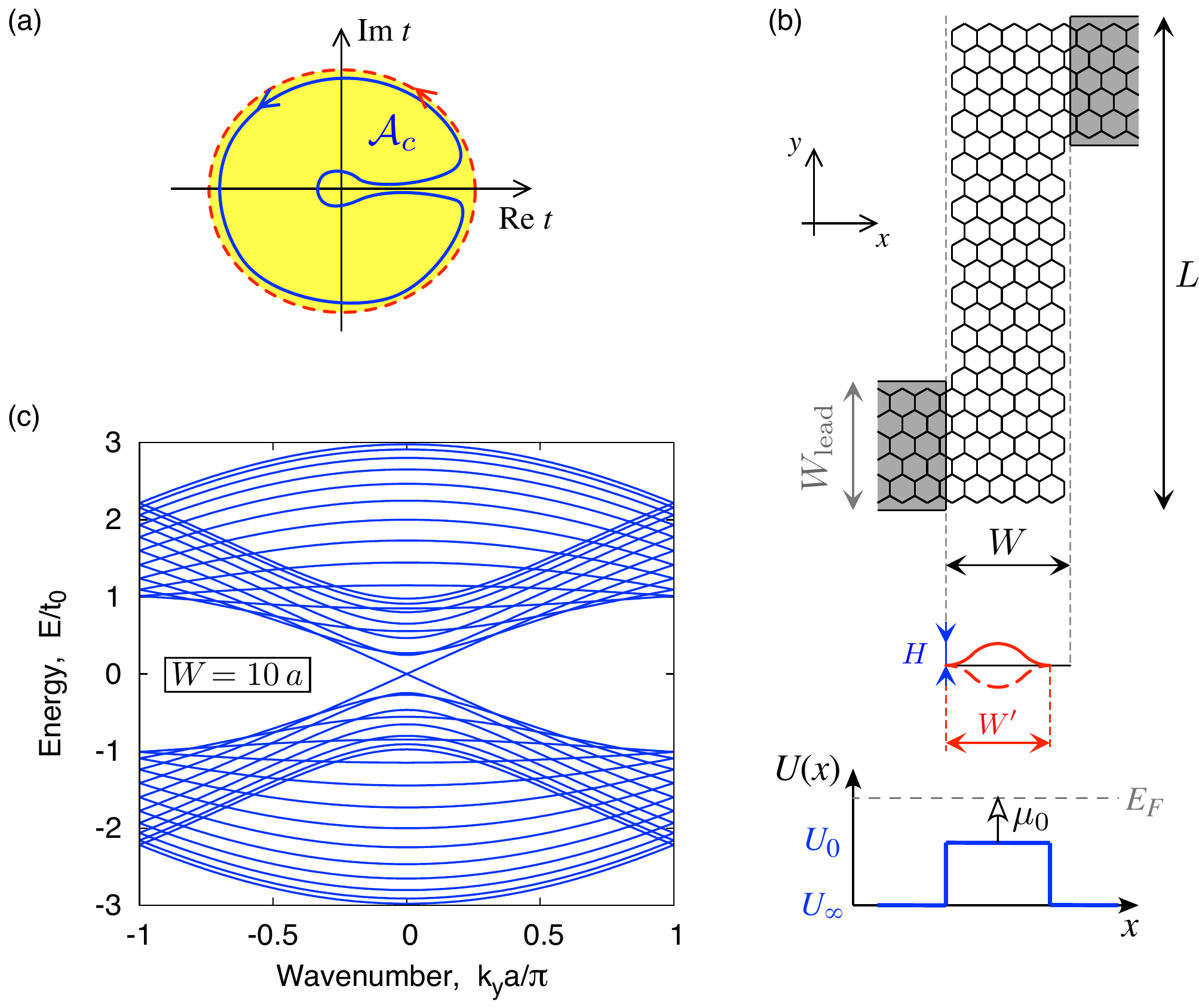}}
\caption{
Buckled graphene nanoribbon as a~quantum pump.
(a) Typical pumping cycles (schematic) for a~single channel
characterized by the complex transmission amplitude, 
$t=\mbox{Re}\,t+i\,\mbox{Im}\,t$. Blue solid line enclosing the area
${\cal A}_c$ shows one-parameter cycle allowed for a~standard gate-driven pump,
red dashed line corresponds to the cycle involving a~shift of a~scatterer. 
(b) Flat ribbon with armchair edges attached to heavily-doped graphene leads
(shaded areas). The ribbon width $W=5\,a$ (with $a=0.246\,$nm the lattice
spacing) and length $L=11.5\sqrt{3}\,a$ are chosen for illustrative purposes
only.
A~cross section of buckled ribbon, characterized by the reduced width
$W'<W$ and the buckle height $H>0$, and the electrostatic potential energy
profile $U(x)$, are also shown.  
(c) Band structure of the infinite metallic-armchair nanoribbon
of $W=10\,a$ width, same as used in the computations. 
}
\label{fig:setup1}
\end{figure}

It is known that quantum pump with a~shifting scatterer may show enhanced
charge transfer per cycle in comparison to a~standard gate-driven pump
\cite{Naz09}, see also Fig.\ \ref{fig:setup1}(a)\footnote{
  In the simplest case of a~one-parameter driven, one-channel pump, 
  the charge per cycle is given by $Q=(e/\pi){\cal A}_c$, with ${\cal A}_c$
  being the $\pm$area enclosed by the contour in the $(\mbox{Re}\,t,
  \mbox{Im}\,t)$ plane (where $t$ is a~parameter-dependent transmission
  amplitude). 
}. 
Motivated by this conjecture we consider a~buckled nanoribbon, see Fig.\
\ref{fig:setup1}(b), similar to the one studied numerically by Yamaletdinov
et.\ al.\ \cite{Yam17,Yam19} as a~physical realization of the classical
$\phi^4$ model and its topological solutions ({\em kinks}) connecting
two distinct ground states.
Our setup is supplemented with two heavily-doped graphene leads, attached
to the clamped edges of a~ribbon, allowing to pass electric current along
the system.

\section{Model and methods}
The analysis starts from the Su-Schrieffer-Heeger (SSH) model for the ribbon,
including the hopping-matrix elements corresponding to the nearest-neighbor
bonds on a~honeycomb lattice \cite{Har93,Ma06,Dre98,Gro18}
\begin{equation}
\label{hamribbon}
{\cal H}_{\rm ribbon} = -t_0 \sum_{\langle{}ij\rangle,s}\exp\left(
    -\beta\frac{\delta{}d_{ij}}{d_0}
  \right)\left(
    c_{i,s}^\dagger{}c_{j,s}+c_{j,s}^\dagger{}c_{i,s}
  \right)
  +\frac{1}{2}K\sum_{\langle{}ij\rangle}\left(\delta{}d_{ij}\right)^2, 
\end{equation}
with a~constrain $\sum_{\langle{}ij\rangle}\delta{}d_{ij}=0$, where
$\delta{d_{ij}}$ is the change in bond length, and
$d_0=a/\sqrt{3}$ is the equilibrium bond length defined via the lattice
spacing $a=0.246\,$nm.
The equilibrium hopping integral $t_0=2.7\,$eV, 
$\beta=3$ is the dimensionless electron-phonon coupling,
and $K\approx{}5000\,$eV/nm$^2$ is the spring constant for a~C-C bond. 
The operator $c_{i,s}^\dagger{}$ (or $c_{i,s}$) creates (or annihilates)
a~$\pi$ electron at the $i$-th lattice site with spin $s$. 

In order to determine the spatial arrangement of carbon atoms  in a~buckled
nanoribbon, $\left\{{\bf R}_j=(x_j,y_j,z_j)\right\}$, we first took
the $\left\{x_j\right\}$ and  $\left\{y_j\right\}$ coordinates same as for
a~flat honeycomb lattice in the equilibrium
and set $\left\{z_j\right\}$ according to
\begin{equation}
\label{kinktanh}
  z = H\tanh\left(\frac{y-y_{0}}{\lambda}\right)
  \sin^2\left(\frac{\pi{}x}{W}\right), 
\end{equation}
representing a~topologically non-trivial solution of the $\phi^4$ model,
with $H$ the buckle height, $W$ the ribbon width, $y_0$ and $\lambda$
the kink position and size (respectively). Next, $x$-coordinates
are rescaled according to $x_j\rightarrow{}x_jW'/W$, with $W'$ adjusted
to satisfy $\sum_{\langle{}ij\rangle}\delta{}d_{ij}=0$ in the
$y_0\rightarrow\pm{}\infty$ ({\em ``no kink''\/}) limit.
In particular, $H=2a$ and $W=10a$ corresponds to $W'/W=0.893$. 
We further fixed the kink size at $\lambda=3a$ (closely resembling
the kink profile obtained from the molecular dynamics in Ref.\ \cite{Yam17}); 
this results in relative bond distorsions not exceeding
$|\delta{d_{ij}}|/{d_0}\leqslant{}0.08$. 
Full optimization of the bond lengths in the SSH Hamiltonian
(\ref{hamribbon}), which may lead to the alternating bond pattern
\cite{Gro18}, is to be discussed elsewhere. 

Heavily-doped graphene leads, $x<0$ or $x>W'$ in Fig.\ \ref{fig:setup1}(b),
are modeled as flat regions ($d_{ij}=d_0$) with the electrostatic potential
energy $U_\infty=-0.5t_0$ (compared to $U_0=0$ in the ribbon), corresponding
to $13$ propagating modes for $W_\infty=20\sqrt{3}a$
and the chemical potential $\mu_0=E_F-U_0=0$.
The scattering problem if solved numerically, for each value of the chemical
potential $\mu_0$ and the kink position $y_0$, using the {\sc Kwant} package
\cite{Kwant} allowing to determine the scattering matrix
\begin{equation}
\label{smatrix}
  S(\mu_0,y_0)=\left(\begin{array}{cc}
    r & t' \\
    t & r' \\
  \end{array}\right), 
\end{equation}
which contains the transmission $t$ ($t'$) and reflection $r$ ($r'$)
amplitudes for charge carriers incident from the left (right) lead,
respectively.

The linear-response conductance can be determined from the $S$-matrix 
via the Landuer-B\"{u}ttiker formula \cite{Naz09}, namely 
\begin{equation}
  G=G_0\mbox{Tr}\,tt^\dagger = \frac{2e^2}{h}\sum_nT_n, 
\end{equation}
where $G_0=2e^2/h$ is the conductance quantum and $T_n$
is the transmission probability for the $n$-th normal mode.
Similarly, in the absence of a~voltage bias, the charge transferred
between the leads upon varying solely the parameter $y_0$ is given by
\begin{equation}
  \label{delq}
  \Delta{}Q=-\frac{ie}{2\pi}\sum_{j}\int{}dy_0\left(
  \frac{\partial{}S}{\partial{}y_0}S^\dagger
  \right)_{jj}, 
\end{equation}
where the summation runs over the modes in a~selected lead.
Additionaly, the integration in Eq.\ (\ref{delq}) is performed
for a~truncated range of 
$-\Lambda\leqslant{}y_0\leqslant{}L+\Lambda$, with
$\Lambda=250\,a\gg\lambda$ for $L=75.5\sqrt{3}\,a$.

\begin{figure}[!t]
\centerline{\includegraphics[width=10.0cm]{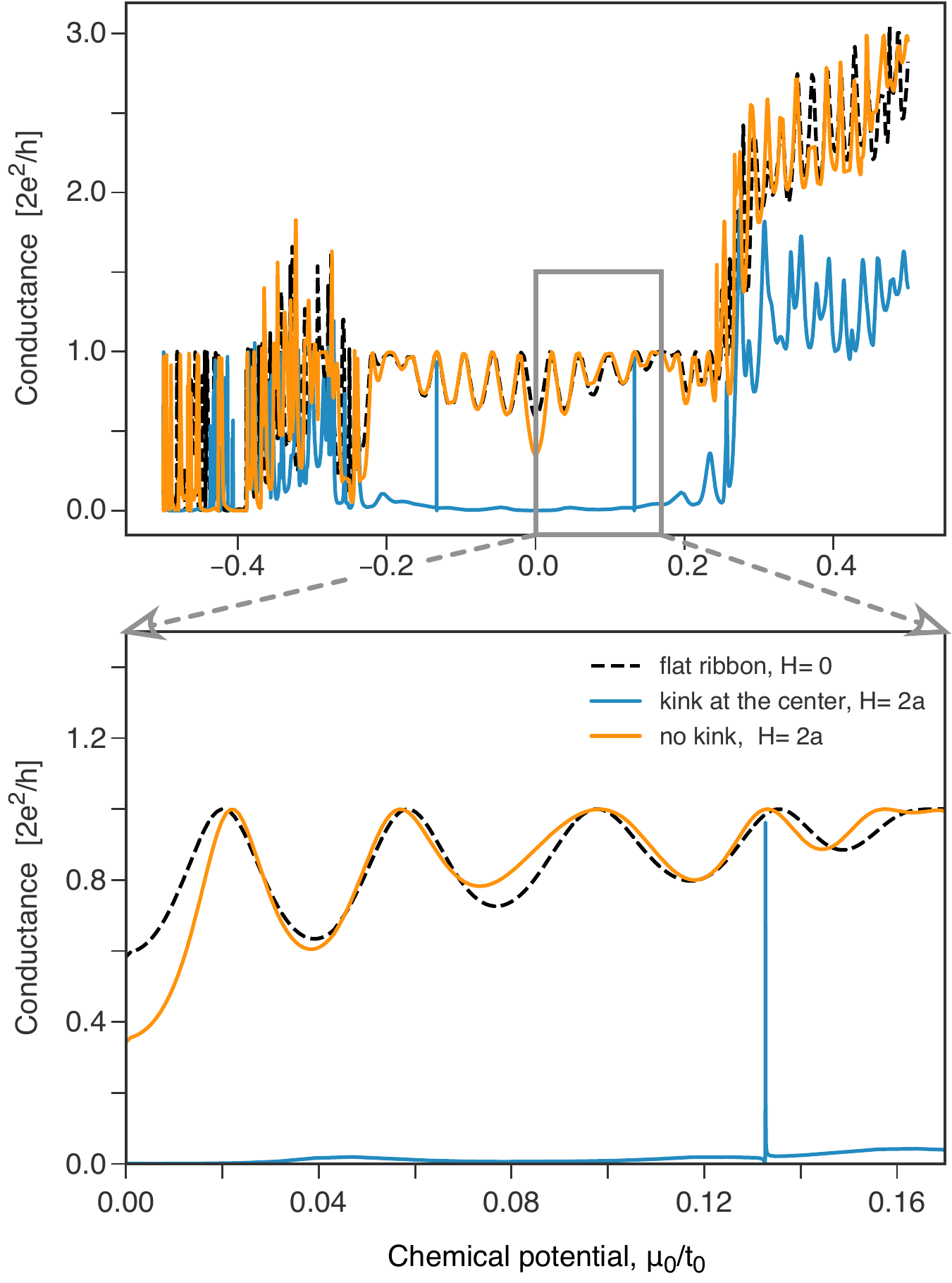}}
\caption{
Conductance of the ribbon with $W=10\,a$ as a~fuction of the chemical
potential. The remaining parameters are $L=75.5\sqrt{3}\,a$ and
$W_{\rm lead}=20\sqrt{3}\,a$. Different lines (same in two panels)
correspond to the flat ribbon geometry, $H=0$, $W'=W$ (black dashed),
the buckled ribbon with $H=2a$ and $y_0=L/2$ 
(blue solid) or $y_0\rightarrow{}-\infty$ (orange solid); 
see Eq.\ (\ref{kinktanh}).
Bottom panel is a~zoom-in of the data shown in top panel. 
}
\label{fig:conductance}
\end{figure}

\section{Results and discussion}
In Fig.\ \ref{fig:setup1}(c) we depict the band structure for an infinite
(and flat) metallic-armchair ribbon of $W=10\,a$ width.
It is remarkable that the second and third lowest-lying subband above the
charge-neutrality point (as well as the corresponding highest subbands
below this point)
show an almost perfect degeneracy near their minima corresponding to
$E_{\rm min}^{(2,3)}\approx{}0.25\,t_0$, which can be attributed to the presence
of two valleys in graphene. For higher subbands, the degeneracy splitting
due to the trigonal warping is better pronounced.

The approximate subband degeneracy has a~consequence for the conductance
spectra of a~finite ribbon, presented in Fig.\ \ref{fig:conductance}:
For both the flat ribbon ($H=0$) and a~buckled ribbon with no kink
($H=2a$, $y_0\rightarrow{}-\infty$) the conductance $G\approx{}G_0$ for
$|\mu_0|<E_{\rm min}^{(2,3)}$ and quickly rises to $G/G_0\approx{}2.5\!-\!3$
for $\mu>E_{\rm min}^{(2,3)}$. For $\mu<0$ the quantization is not so apparent,
partly due the presence of two p-n interfaces (at $x=0$ and $x=W'$) leading to
stronger-pronounced oscillations (of the Fabry-Perrot type), and partly due to
a~limited number of propagating modes in the leads. 
For a~kink positioned at the center of the ribbon ($H=2a$, $y_0=L/2$)
the conductance is strongly suppressed, $G\ll{}G_0$, in the full
$|\mu_0|<E_{\rm min}^{(2,3)}$ range (with the exception from two narrow
resonances at $\mu_0=\pm{}0.1327\,t_0$, which vanishes for $y_0\neq{}L/2$)
defining a~feasible energy window for the pump operation.

\begin{figure}[!t]
\centerline{\includegraphics[width=10.0cm]{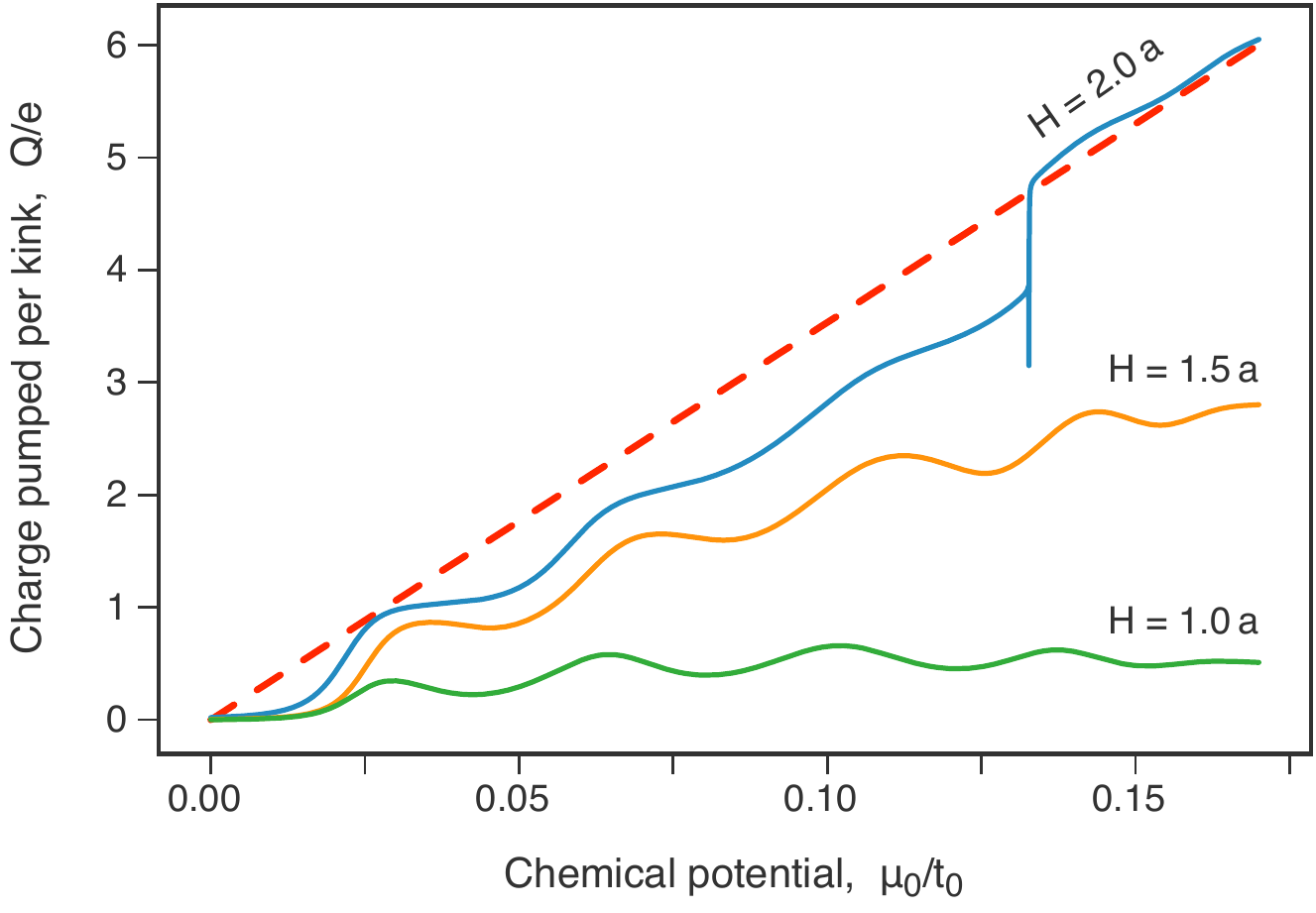}}
\caption{ 
  Charge transferred upon adiabatic kink transition calculated from
  Eq.\ (\ref{delq}), for varying buckling amplitude $H/a=1,\,2$ 
  (specified for each line in the plot), displayed as a~function of
  the chemical potential. 
  Dashed line shows the approximation given by Eq.\ (\ref{qtotest}). 
  Remaining parameters are same as in Fig.\ \ref{fig:conductance}. 
}
\label{fig:pumping}
\end{figure}

In Fig.\ \ref{fig:pumping} we display the charge transferred between
the leads at zero bias, see Eq.\ (\ref{delq}), when slowly moving a~kink
along the ribbon ({\em adiabatic kink transition}).
In the energy window considered, the ribbon dispersion relation consists
of two subbands with opposite group velocities,
$E_{\pm}^{(1)}=\pm{}\hbar{}v_Fk_y$, see Fig.\ \ref{fig:setup1}(c).
Therefore, the total charge available for
transfer at the ribbon section of $L_{\rm eff}=L-W_{\rm infty}$ length
can be estimated (up to a~sign) as
\begin{equation}
\label{qtotest}
  \frac{Q_{\rm tot}}{e}\approx\frac{|\mu_0|}{\hbar{}v_F}\frac{L_{\rm eff}}{\pi}
  = 35.3\,\frac{|\mu_0|}{t_0},
\end{equation}
where $v_F=\sqrt{3}\,t_0{}a/(2\hbar)\approx{}10^6\,$m/s is the Fermi
velocity in graphene and we put $L\!-\!W_{\rm infty}=55.5\sqrt{3}\,a$. 
For the strongest deformation ($H=2a$) the kink almost perfectly blocks the
current flow, and the charge transferred is close to the approximation
given by Eq.\ (\ref{qtotest}); see Fig.\ \ref{fig:pumping}.

\section{Conclusion}
We have investigated, by means of computer simulations of electron transport,
the operation of adiabatic quantum pump build within a~topological defect
moving in buckled graphene nanoribbon.
We find that 
the pump effectiveness is close to the maximal (corresponding to a~kink
perfectly blocking the current flow) for moderate bucklings, with 
relative bond distortions not exceeding $8\%$.
As topological defects generally move with negligible energy dissipation,
we hope our discussion will be a~starting point to design new graphene-based
energy conversion devices.

\section*{Acknowledgments}
We thank Tomasz Roma\'{n}czukiewicz for discussions.
The work was supported by the National Science Centre of Poland (NCN)
via Grant No.\ 2014/14/E/ST3/00256.


\end{document}